\documentclass[12pt,preprint]{emulateapj}
\usepackage{natbib}
\usepackage{aas_macros}

\DeclareMathAlphabet{\mathsc}{OT1}{cmr}{m}{sc}

\begin{document}

\title{
Evolution of the Fraction of Clumpy Galaxies at $0.2<z<1.0$\\ in the COSMOS field\altaffilmark{*}}

\author{ 
K.~L.~Murata\altaffilmark{1,2},
M.~Kajisawa\altaffilmark{2,3},
Y.~Taniguchi\altaffilmark{3},
M.~A.~R. Kobayashi\altaffilmark{3},
Y. Shioya\altaffilmark{3},
P.~Capak\altaffilmark{4},
O.~Ilbert\altaffilmark{5},
A.~M.~Koekemoer\altaffilmark{6},
M.~Salvato\altaffilmark{7}, 
N.~Z.~Scoville\altaffilmark{8}
}

\altaffiltext{*}{Based on observations with the NASA/ESA 
        {\it Hubble Space Telescope}, obtained at the Space Telescope Science 
	Institute, which is operated by AURA Inc, under NASA contract NAS 
	5-26555. Also based on observations made with the Spitzer Space
	Telescope, which is operated by the 
	Jet Propulsion Laboratory, California Institute of Technology, 
	under NASA contract 1407. Also based on data collected at;  
	the Subaru Telescope, which is operated by the National Astronomical 
	Observatory of Japan; the XMM-Newton, an ESA science mission with 
	instruments and contributions directly funded by ESA Member States and
	NASA; the European Southern Observatory under Large 
	Program 175.A-0839, Chile; Kitt Peak National Observatory, Cerro 
	Tololo Inter-American Observatory and the National Optical Astronomy 
	Observatory, which are operated by the Association of Universities for 
	Research in Astronomy, Inc. (AURA) under cooperative agreement with 
	the National Science Foundation; and the Canada-France-Hawaii 
	Telescope with MegaPrime/MegaCam operated as a joint project by the 
	CFHT Corporation, CEA/DAPNIA, the NRC and CADC of Canada, the CNRS 
	of France, TERAPIX and the Univ. of Hawaii.}

\altaffiltext{1}{Department of Particle and Astrophysical Science, Nagoya University, 
Furo-cho, Chikusa-ku, Nagoya 464-8602, Japan
{\it e-mail murata.katsuhiro@g.mbox.nagoya-u.ac.jp}
}
\altaffiltext{2}{Graduate School of Science and Engineering, Ehime University, 
        Bunkyo-cho, Matsuyama 790-8577, Japan}
\altaffiltext{3}{Research Center for Space and Cosmic Evolution, 
        Ehime University, Bunkyo-cho, Matsuyama 790-8577, Japan}
\altaffiltext{4}{Spitzer Science Center, 314-6 Caltech, Pasadena, CA 91125, USA}
\altaffiltext{5}{Aix Marseille Universit\'e, CNRS, LAM (Laboratoire d'Astrophysique de Marseille), UMR 7326, 13388, Marseille, France}
\altaffiltext{6}{Space Telescope Science Institute, 3700 San Martin Drive, Baltimore, MD 21218, USA}
\altaffiltext{7}{Max Planck Institut f\"{u}r Plasma Physik and Excellence Cluster, 85748 Garching, Germany}
\altaffiltext{8}{California Institute of Technology, MC 249-17, 1200 East California Boulevard, Pasadena, CA 91125, USA}

\shortauthors{K. Murata et al.}
\shorttitle{Clumpy Galaxies at $0.2<z<1.0$ in COSMOS}

\begin{abstract}
Using the {\it Hubble Space Telescope}/Advanced Camera for Surveys 
data in the COSMOS field, we systematically searched 
clumpy galaxies at $0.2<z<1.0$ and investigated the fraction of clumpy galaxies 
and its evolution as a function of stellar mass, star formation rate 
(SFR), and specific SFR (SSFR).
The fraction of clumpy galaxies in star-forming galaxies with $M_{\rm star} > 10^{9.5} 
M_{\odot}$ decreases with time from $\sim 0.35$ at $0.8<z<1.0$ to $\sim 0.05$ 
at $0.2<z<0.4$ irrespective of the stellar mass, 
although the fraction tends to be slightly lower 
for massive galaxies with $M_{\rm star} > 10^{10.5} M_{\odot}$ at each redshift.
On the other hand, the fraction of clumpy galaxies increases with increasing both 
SFR and SSFR in all the 
redshift ranges we investigated. In particular, we found that 
the SSFR dependences of the fractions are similar among galaxies with different 
stellar masses, and the fraction at a given 
SSFR does not depend on the stellar mass in each redshift bin.
The evolution of the fraction of clumpy galaxies from $z\sim 0.9$ to $z\sim0.3$ 
seems to be explained 
by such SSFR dependence of the fraction and the evolution of SSFRs of 
star-forming galaxies. 
The fraction at a given SSFR also appears to decrease with time, 
but this can be due to the effect of the morphological K-correction. 
We suggest that these results are understood by the gravitational fragmentation model 
for the formation of giant clumps in disk galaxies, where the gas mass fraction is 
a crucial parameter.

\end{abstract}



\section{INTRODUCTION}
In the present universe, most bright galaxies have regular and symmetric morphologies, 
which can be classified in the framework of the Hubble sequence \citep{hub36}.
On the other hand, using 
the high-resolution imaging capability of the {\it Hubble Space Telescope}
 ($HST$), it has been found 
that many star-forming galaxies at $z>1$ have irregular shapes 
 with asymmetric structures,  
(e.g., \citealp{cow95}; \citealp{ste96}; \citealp{kaj01}; \citealp{elm07};
\citealp{cam11}). Although these high-redshift irregular galaxies show a variety of 
morphologies, they commonly have giant (kpc scale) star-forming clumps (e.g.,
 \citealp{elm09a}; \citealp{for11}). Recent NIR integral field spectroscopy 
observations of star-forming clumpy galaxies at $z\sim2$ revealed that 
a significant fraction of these galaxies show coherent rotation with a relatively 
large turbulent velocity in their ionized gas kinematics (e.g., \citealp{for06};
 \citealp{wri07}; \citealp{gen08}; \citealp{cre09}; \citealp{for09}).
Several studies of the radio CO line observations also found that 
actively star-forming galaxies at $1\lesssim z \lesssim 3$ have large 
gas mass fractions of $\sim 0.3$ -- 0.8 (\citealp{dad10}; \citealp{tac10}; \citealp{tac13}).
While some of these galaxies are  
 galaxy mergers (e.g., \citealp{som01}; \citealp{lot04}; \citealp{pue10}), 
these results can be 
explained by theoretical models where gas-rich rotational disks are gravitationally 
unstable for the fragmentation and lead to the formation of giant star-forming clumps
(e.g., \citealp{nog98}; \citealp{imm04}; \citealp{bou07}; \citealp{dek09}). 
The high gas mass fraction of these galaxies is considered to be maintained by 
the rapid and smooth cosmic infall of gas along large-scale filaments. 
 Since the accretion rate of gas is expected to decrease with time, 
especially at $z\lesssim 1$, 
the gas fraction of these clumpy galaxies declines at lower redshifts 
as the gas consumption by the star formation proceeds, which results in 
the stabilization of the gas disks \citep{cac12}.
In this view, these high-redshift clumpy galaxies are considered to be 
progenitors of normal 
(disk) galaxies at low redshifts. 
Therefore, it is important to study the evolution of these clumpy galaxies 
in order to understand the formation process of normal galaxies in the present 
universe. 

However, the number of systematic surveys for clumpy galaxies is very limited, 
because wide-field imaging data with high spatial resolution 
are required. While \cite{elm07} claimed that clumpy galaxies are dominated 
at high redshift based on the morphological analysis of galaxies in the HUDF field, 
\cite{tad14} reported that $\sim $40\% of 100 H$\alpha$ emitters at $z\sim 2.2$ and 
$z\sim 2.5$ in the UKIDSS/UDS-CANDELS field show clumpy morphologies.
\cite{wuy12} measured the fraction of clumpy galaxies in star-forming galaxies 
with $M_{\rm star} > 10^{10} M_{\odot}$ at $z\sim2$ in the GOODS-South field, and  
found that the fraction is 74\% for clumps selected at the rest-frame 2800 \AA\ and 
42\% for those selected at the rest-frame $V$ band, which suggests that  
the morphological K-correction can be important for the selection of clumpy galaxies.
Although systematic surveys of clumpy galaxies at lower redshifts are also important for  
understanding the connection between clumpy galaxies at high redshifts and normal galaxies 
in the nearby universe, there is few survey for clumpy galaxies at $z\lesssim 1$. 
In this paper, we systematically search clumpy galaxies at $0.2<z<1.0$ in the 
COSMOS field \citep{sco07} and investigate their physical properties.
The high spatial resolution images taken with $HST$/Advanced Camera for Surveys 
(ACS) over the very wide field 
allow us to construct a large sample of clumpy galaxies at $z<1$ 
and to investigate the fraction of clumpy galaxies and its evolution as a function of 
physical properties such as stellar mass and star formation rate 
 for the first time.
Section \ref{sec:ana} describes our sample and details of the selection method
for clumpy galaxies. We present the physical properties of clumpy galaxies and 
investigate the fraction of these galaxies and its evolution in Section \ref{sec:result}.
 In Section \ref{sec:dis}, we summarize our results and discuss their implications. 
Throughout this paper, magnitudes are given in the AB system. We adopt a flat universe 
with $\Omega_{\rm matter}=0.3$, $\Omega_{\Lambda}=0.7$, and $H_{0}=70$ km 
s$^{-1}$ Mpc$^{-1}$.

\begin{figure*}
\begin{center}
\includegraphics[width=180mm]{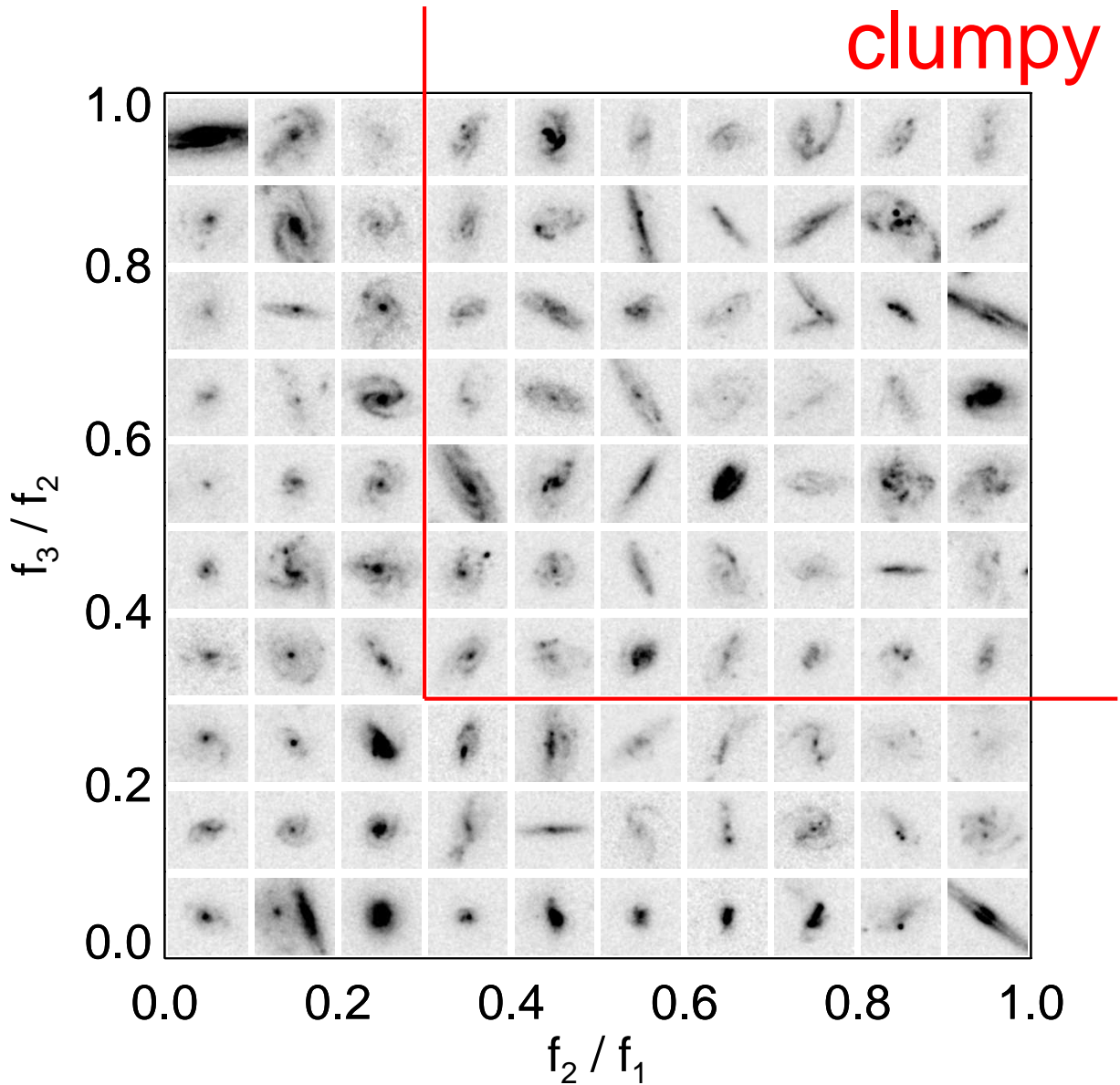}
\caption{
$HST$/ACS $I_{\rm F814W}$-band images of galaxies with more than two components 
as a function of the flux ratios among the brightest three clumps in each galaxy.
$f_{\rm 2}/f_{\rm 1}$ is the ratio between the second brightest and the brightest 
clumps, while $f_{\rm 3}/f_{\rm 2}$ is that between the third and second brightest 
clumps. The red line shows the criteria for clumpy galaxies ($f_{\rm 2}/f_{\rm 1} \ge 
0.3$ \& $f_{\rm 3}/f_{\rm 2} \ge 0.3$). 
These galaxies are randomly selected from those with each range of the parameters.
}
\label{fig:monclump}
\end{center}
\end{figure*}

\begin{figure*}
\begin{center}
\includegraphics[width=170mm]{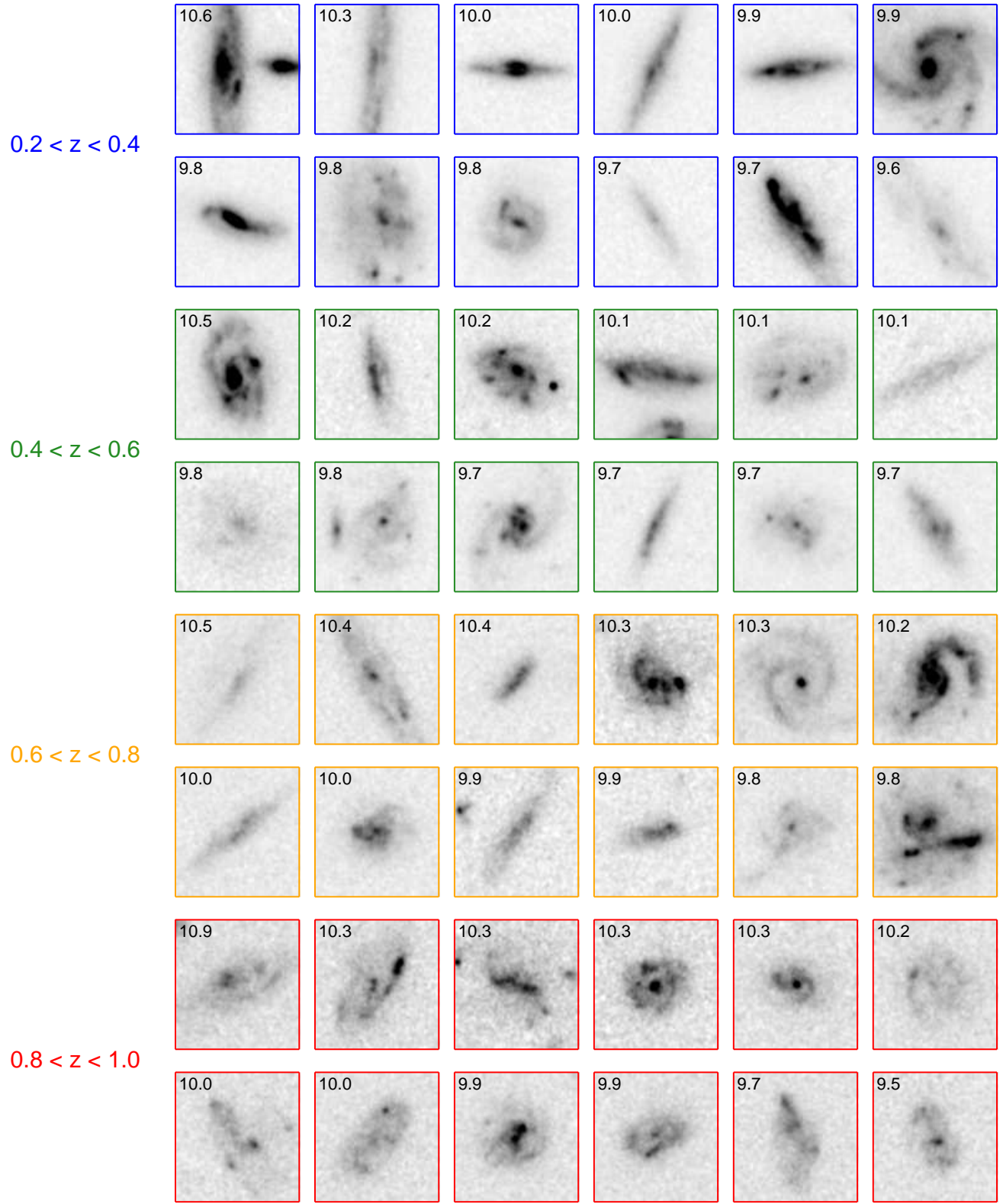}
\caption{
Examples of clumpy galaxies in the redshift bins. 
Galaxies are randomly selected in each redshift bin and 
are shown in the order of their stellar mass.
The number in each panel shows $\log(M_{\rm star}/M_{\odot})$. 
}
\label{fig:clumpy}
\end{center}
\end{figure*}

\begin{figure*}
\begin{center}
\includegraphics[width=170mm]{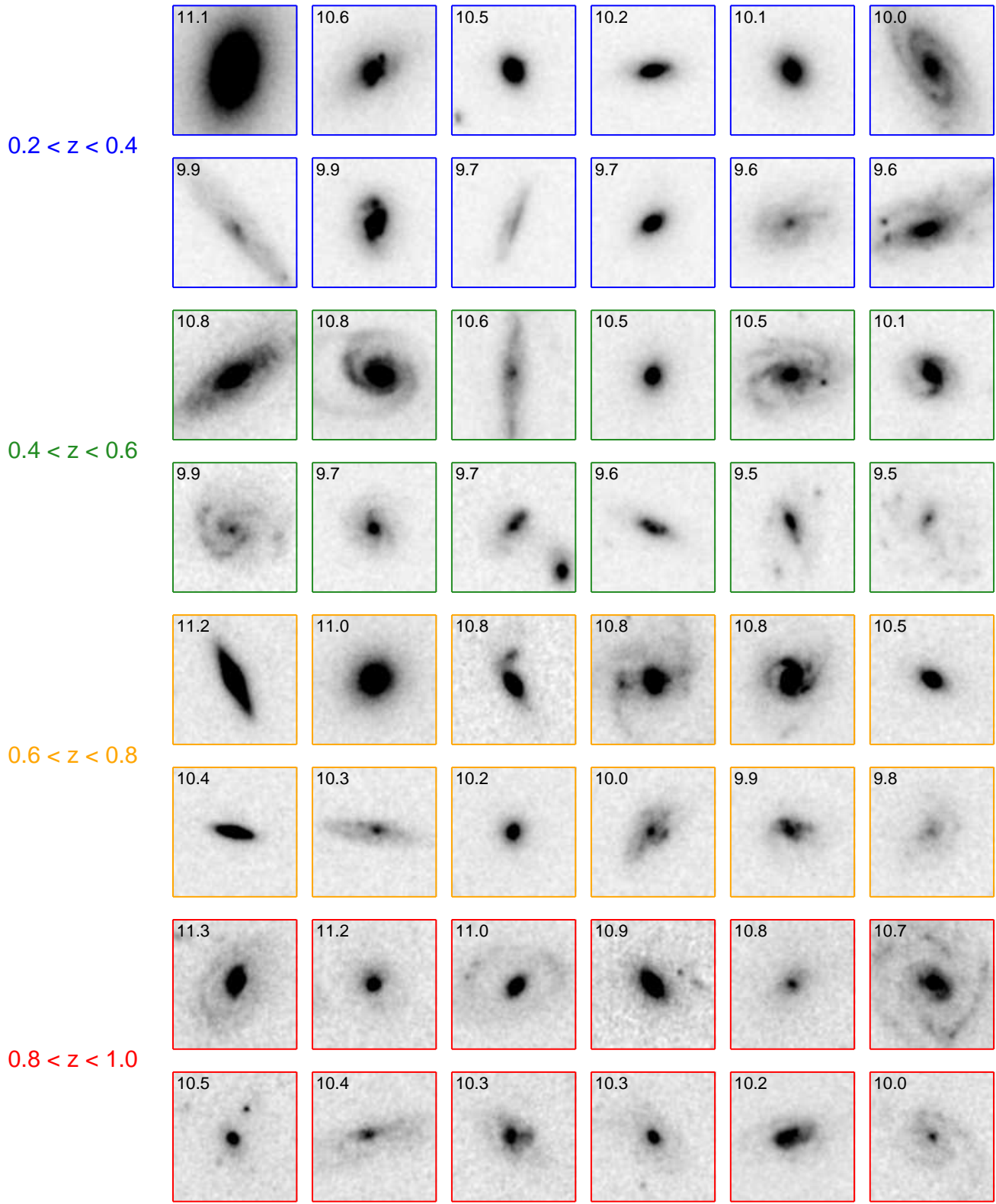}
\caption{
The same as Figure \ref{fig:clumpy} but for non-clumpy galaxies.
}
\label{fig:nonclumpy}
\end{center}
\end{figure*}

\begin{figure*}
\begin{center}
\includegraphics[width=50mm,angle=-90]{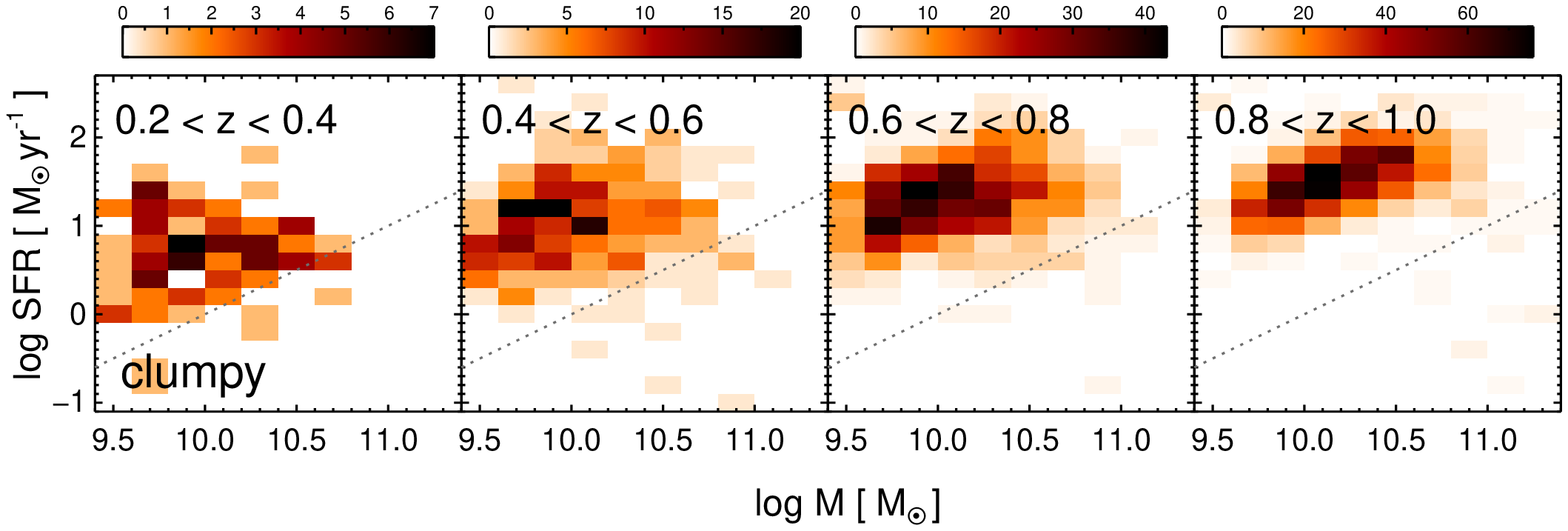}
\includegraphics[width=50mm,angle=-90]{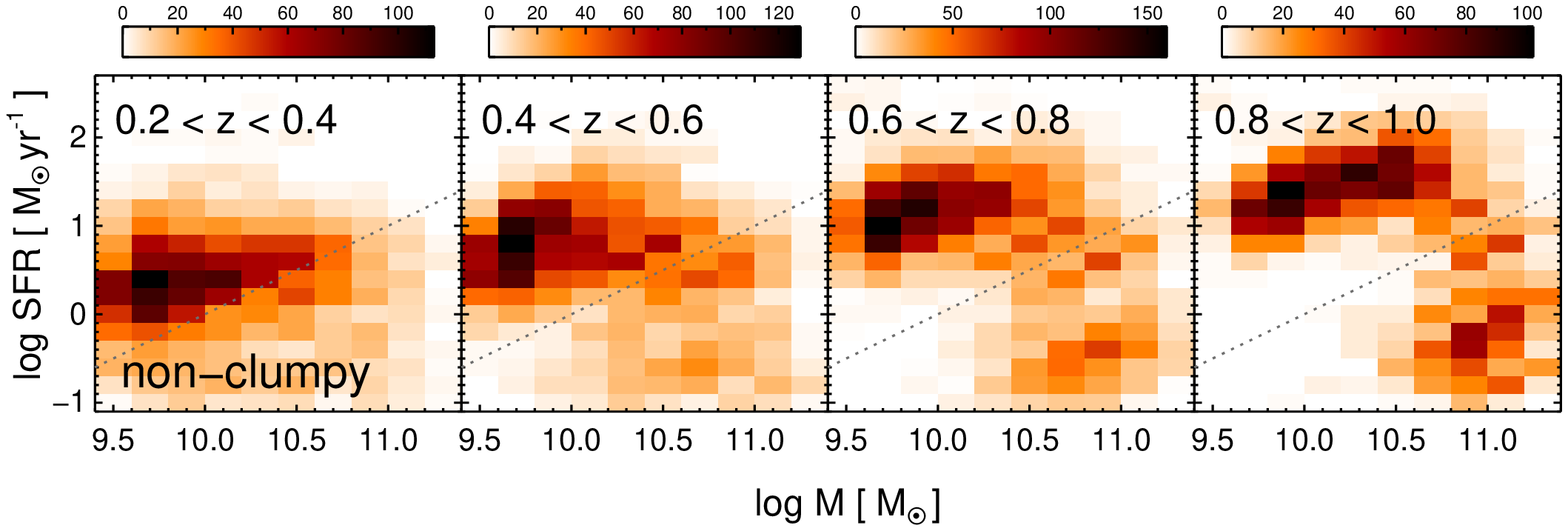}
\caption{
SFR vs. stellar mass for clumpy galaxies (top panels) and non-clumpy galaxies 
(bottom panels) in each redshift bin. The dashed line shows a constant SSFR of 
0.1 Gyr$^{-1}$, above which galaxies are classified as star-forming ones.
}
\label{fig:mstar}
\end{center}
\end{figure*}

\section{SAMPLE AND ANALYSIS}
\label{sec:ana}

\subsection{Sample}
In this study, we used a sample of galaxies with photometric redshifts of 
$0.2<z<1.0$ from the COSMOS photometric redshift catalog \citep{ilb09}.
We basically analyzed galaxies with $I_{\rm F814W} < 22.5$ in order to securely 
select clumpy galaxies in the $HST$/ACS $I_{\rm F814W}$-band images (see the 
next subsection). The photometric redshift is estimated with more than 30 bands 
 data from UV to MIR wavelength in the COSMOS field, and its uncertainty 
is very small ($\Delta z \lesssim 0.02$) for galaxies with $I_{\rm F814W} < 22.5$ 
at $z<1$ \citep{ilb09}, which is sufficiently accurate for our purpose. 

In order to investigate the fraction of clumpy galaxies as a function of 
the physical properties of galaxies,  
we used the stellar mass (M$_{\rm star}$) and the star formation rate (SFR) 
of our sample galaxies 
estimated from the spectral energy distribution 
(SED) fitting technique (\citealp{ilb10}; \citealp{ilb13}).
The multi-band photometric data from UV to MIR wavelength were fitted with 
the population synthesis model by \cite{bru03}. The 
exponentially declining star formation histories and the \cite{cal00} extinction 
law were assumed, and the \cite{cha03} IMF was adopted (see \citealp{ilb13} for 
details). The best-fit model was used 
to convert the luminosities to the stellar mass and the SFR.
We excluded X-ray sources detected in the {\it Chandra} or {\it XMM-Newton} images 
(\citealp{has07}; \citealp{elv09}) and galactic stars classified in the SED fitting 
from our sample.  

\begin{table}
\begin{center}
\caption{Number of galaxies in our sample}
\label{tab:sample}

\begin{tabular}{ccc}
\hline\hline
Redshift & All ($M_{\rm star}>10^{9.5}M_{\odot}$) & Clumpy ($M_{\rm star}>10^{9.5}M_{\odot}$)\\
\hline
$0.2 \le z < 0.4$ & 7392 (3826) & 363 (106) \\
$0.4 \le z < 0.6$ & 5742 (4221) & 464 (304) \\
$0.6 \le z < 0.8$ & 6297 (5597) & 895 (781) \\
$0.8 \le z < 1.0$ & 4596 (4500) & 1081 (1055) \\ 
\hline
total & 24027 (18144) & 2803 (2246) \\
\hline
\end{tabular}
\end{center}
\end{table}

\subsection{Selection for Clumpy Galaxies}
Using the 
$HST$/ACS $I_{\rm F814W}$-band data of the COSMOS survey 
(version 2.0, \citealp{koe07}), 
we  examined the morphology of our sample galaxies 
and selected clumpy galaxies quantitatively.
The pixel scale of the reduced ACS images is 0.03 arcsec and the FWHM of the Point 
Spread Function is $\sim 0.1$ arcsec.  
The clumpy galaxy is characterized by several relatively bright components (clumps) 
in a galaxy. We here consider galaxies with more than two such clumps 
($N_{\rm clump} \ge 3$) as clumpy galaxies.
In order to select clumpy galaxies, we first detected galaxies on the $I_{\rm F814W}$-band
data, using the SExtractor software \citep{ber96}. A detection threshold of 
2.0 times the local background root mean square over 15 connected pixels was used.
We adopted the {\it DEBREND\_NTHRESH} parameter of 64 and the {\it DEBLEND\_MINCONT} 
parameter of 0.1 in the first SExtractor run. We then searched counterparts of 
our sample galaxies mentioned above on the $I_{\rm F814W}$-band image, 
and found the counterparts for 24027 galaxies out of 24176 galaxies 
with $I_{\rm F814W} < 22.5$ at $0.2<z<1.0$. 
Next, we changed the {\it DEBLEND\_MINCONT} 
parameter to 0.001 and rerun the SExtractor to more aggressively deblend each galaxy 
and detect relatively bright clumps in the galaxy.
The resulting catalog in this second SExtractor run was cross-matched
with that in the first run. 
We then selected sample galaxies which are deblended into more than 
two components as clumpy galaxy candidates. 
In order to ensure that at least three clumps are comparatively bright, 
we set further criteria for clumpy galaxies as 
\begin{equation}
f_{\rm 2}/f_{\rm 1} \ge 0.3,
\end{equation}
and
\begin{equation}
f_{\rm 3}/f_{\rm 2} \ge 0.3, 
\end{equation}
where $f_{\rm 1}$, $f_{\rm 2}$, and $f_{\rm 3}$ are $I_{\rm F814W}$-band 
fluxes of the brightest, 
the second brightest, and the third brightest clumps, respectively, in the galaxy.
In Figure \ref{fig:monclump}, we show examples of galaxies 
with more than two clumps as a function of these flux ratios among the brightest 
three clumps to demonstrate our classification. 
One can see that our selection enable to pick up sources with 
several significant clumps. On the other hand, galaxies with one dominant 
component, which corresponds to a bulge in some cases, lie at the left side
in the figure, while those with only two bright components are located at the 
bottom right. In this paper, we refer to all galaxies which are not satisfied by 
the criteria for clumpy galaxies as ``non-clumpy'' galaxies.  

Figures \ref{fig:clumpy} and \ref{fig:nonclumpy} show examples of clumpy and 
non-clumpy galaxies in each redshift bin, respectively. 
We selected total 2803 clumpy galaxies at $0.2<z<1.0$ with $I_{\rm F814W} < 22.5$.
Our sample sizes are summarized in Table \ref{tab:sample}.


\section{RESULTS}
\label{sec:result}
\subsection{Stellar Mass and SFR of Clumpy Galaxies} 
We show the M$_{\rm star}$--SFR diagram for the clumpy  
and non-clumpy galaxies, respectively, in Figure \ref{fig:mstar}.
Relatively small number of galaxies around $M_{\rm star} \sim 10^{9.5} M_{\odot}$ 
in the both samples is due to the magnitude limit of $I_{\rm F814W}<22.5$. 
Since the observed $I_{\rm F814W}$ band samples a shorter rest-frame wavelength 
at higher redshift, low-mass galaxies with lower SFRs in the high redshift bins  
tend to be missed by this magnitude limit. We discuss possible systematic effects 
of the magnitude limit on our results in Section \ref{sec:bias}.

The distribution of non-clumpy galaxies in the M$_{\rm star}$--SFR diagram 
shows a bimodality, which consists of passively evolving galaxies mainly located 
at higher stellar mass around $M_{\rm star} \sim 10^{11} M_{\odot}$ and 
star-forming galaxies located at $M_{\rm star} \lesssim 10^{10.5} M_{\odot}$.
The SFRs of star-forming galaxies increase with increasing stellar mass, and they 
 form a sequence in the M$_{\rm star}$--SFR plane, namely, the ``main sequence''
of star-forming galaxies (e.g., \citealp{noe07}). The SFRs of star-forming galaxies 
at a given stellar mass increase with increasing redshift over $0.2<z<1.0$.
Such distribution of galaxies in the M$_{\rm star}$--SFR plane 
and its evolution at $z\lesssim1$ are consistent with previous studies 
(e.g., \citealp{noe07}; \citealp{san09}; \citealp{kaj10}).

On the other hand, the clumpy galaxies are preferentially located on the main sequence 
of star-forming galaxies. In Figure \ref{fig:mstar}, 
we plot the boundary line of the $SFR/M_{\rm star} = 0.1$  
 Gyr$^{-1}$, which divides galaxies 
into the passively evolving and star-forming populations, for reference. 
Since almost all clumpy galaxies lie above the boundary line, especially,  
at high redshift, they are star-forming galaxies. 
The range of the SFRs of clumpy galaxies at a given stellar mass is similar 
to that of the other star-forming galaxies in all the mass and redshift ranges, 
although their SFRs tend to be higher values 
as we will show in detail in the following subsection.

\subsection{Fraction of Clumpy Galaxies as a Function of Physical Properties}
\label{sec:frac}
\begin{figure*}
\begin{center}
\includegraphics[width=90mm]{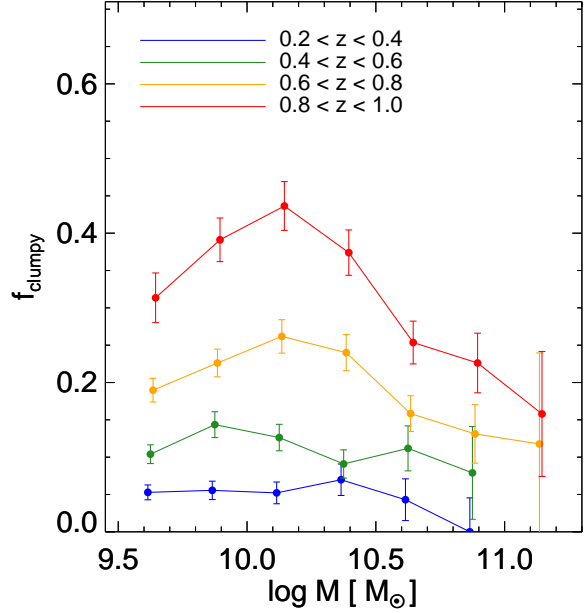} 
\caption{
Fraction of clumpy galaxies in star-forming galaxies with $SSFR > 0.1$ Gyr$^{-1}$ 
as a function of stellar mass for the different redshift bins. The error bars 
are based on the Poisson statistics.
}
\label{fig:mfrac}
\end{center}
\end{figure*}

We investigated the fraction of clumpy galaxies in our sample 
at $0.2<z<1.0$ as a function of 
stellar mass, SFR, and specific SFR ($=$ SFR/M$_{\rm star}$, hereafter SSFR), to 
study their role in the galaxy evolution and the origins of their morphology.
Figure \ref{fig:mfrac} shows the fraction of clumpy galaxies in star-forming galaxies 
with $SSFR > 0.1$ Gyr$^{-1}$ as a function of stellar mass for the different redshift 
bins. The fraction does not strongly depend on stellar mass in all the redshift bins, 
although it tends to be slightly lower at $M_{\rm star} > 10^{10.5} M_{\odot}$. 
We also note that the fraction at $0.8<z<1.0$ becomes slightly higher around 
$M_{\rm star} \sim 10^{10} M_{\odot}$.
In Figure \ref{fig:mstar} we can see a concentration of clumpy galaxies 
at $M_{\rm star} \sim 10^{10} M_{\odot}$ and $SFR \sim 10^{1.75} M_{\odot}$ yr$^{-1}$, 
and therefore some fluctuation in the number density of such galaxies 
in our survey field may cause such high fraction of clumpy galaxies around 
 $M_{\rm star} \sim 10^{10} M_{\odot}$ at $0.8<z<1.0$. 
There may also be the effect of the bias for galaxies with relatively high SSFR 
near the limiting mass at high redshift caused by the magnitude limit of 
$I_{\rm F814W}<22.5$, as we show in Section \ref{sec:bias}.
The fraction of clumpy galaxies decreases with time from $\sim 0.35$ at $0.8<z<1.0$ 
to $\sim 0.05$ at $0.2<z<0.4$. This is consistent with the results in the 
previous studies that bright clumpy galaxies are rare in the present universe, 
while many such galaxies have been observed at $z\gtrsim 1$ (e.g., \citealp{elm07}; 
\citealp{elm09b}).

\begin{figure*}
\begin{center}
\includegraphics[width=90mm]{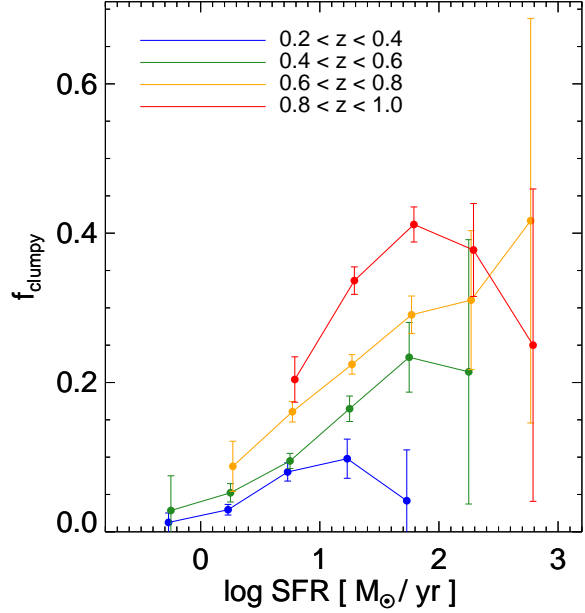} 
\caption{
Fraction of clumpy galaxies in star-forming galaxies with $M_{\rm star} > 10^{9.5} 
M_{\odot}$ as a function of SFR for the different redshift bins.
The error bars 
are based on the Poisson statistics.
}
\label{fig:sfrac}
\end{center}
\end{figure*}

\begin{figure*}
\begin{center}
\includegraphics[width=170mm]{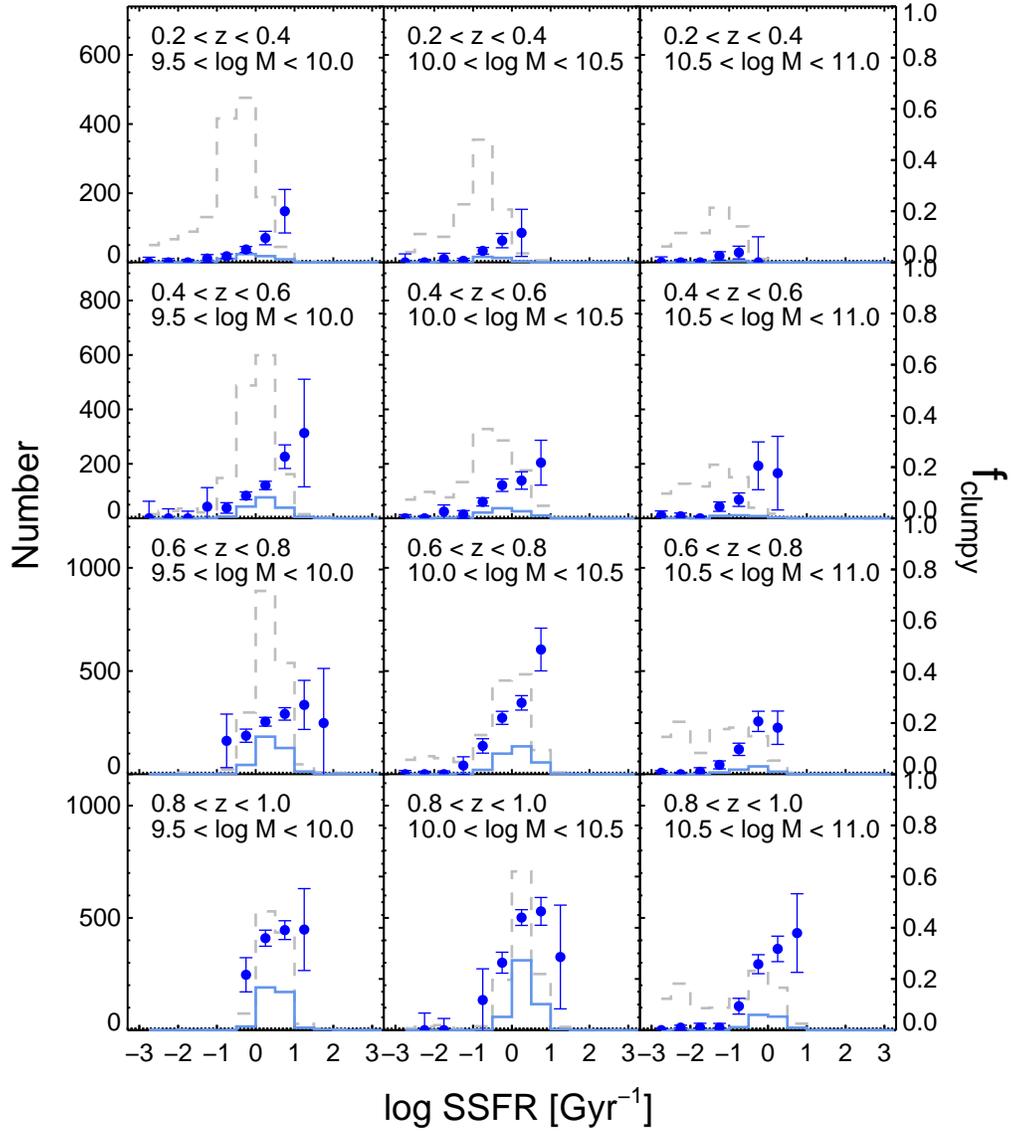} 
\caption{
SSFR distribution for clumpy and non-clumpy galaxies 
(histograms) for each redshift and stellar mass bin. 
The fraction of clumpy galaxies is also shown as a function of SSFR (solid circles, 
right ordinate). 
The solid histogram shows clumpy galaxies, 
while the dashed histogram represents non-clumpy galaxies. 
The error bars are based on the Poisson statistics.
}
\label{fig:msfrac}
\end{center}
\end{figure*}

\begin{figure*}
\begin{center}
\includegraphics[width=150mm]{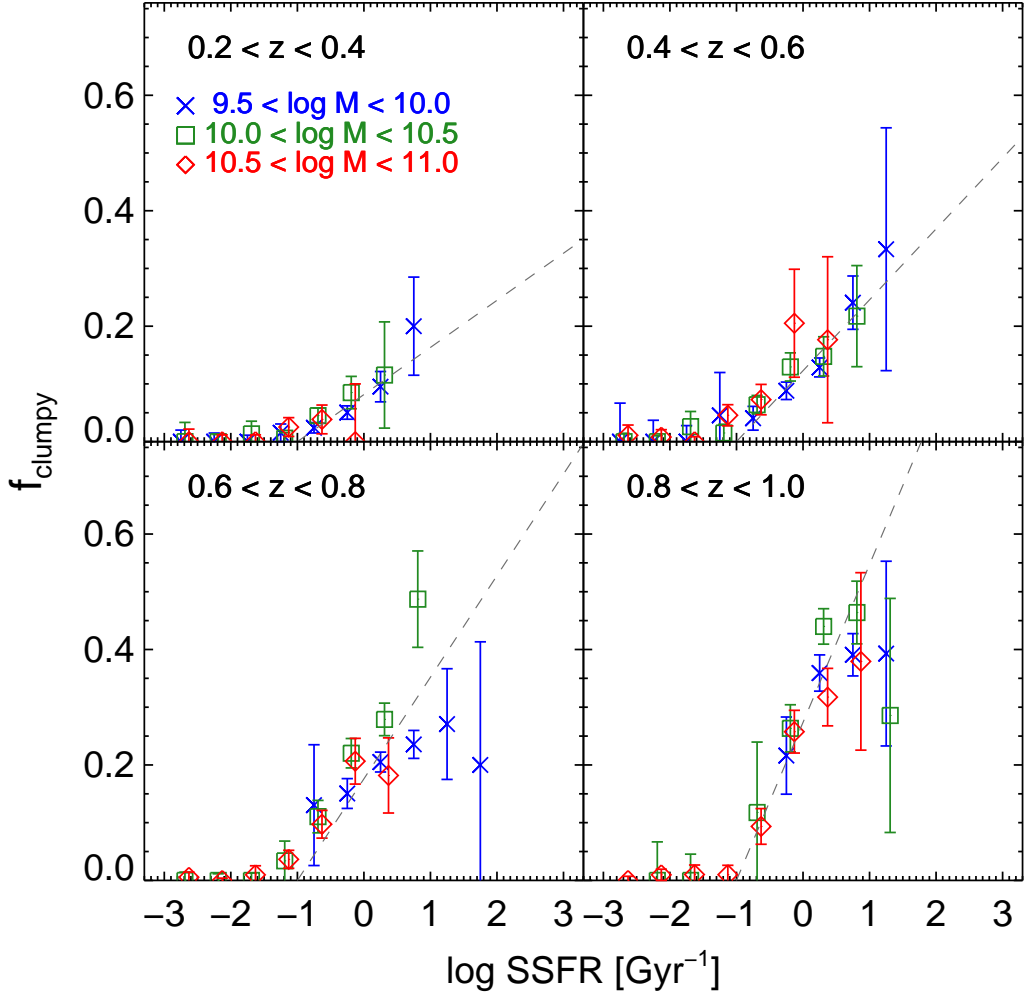} 
\caption{
Fraction of clumpy galaxies as a function of SSFR for each redshift bin.
The fractions for the different stellar mass ranges are shown in the same panel.
The dashed line represents the fitting result of the data points at $SSFR > 0.1$ Gyr$^{-1}$ 
with a linear line of $f_{\rm clumpy} = a \times [ \log(SSFR) + 1.0 ]$, 
where $a$ is a free parameter. 
}
\label{fig:fssfr}
\end{center}
\end{figure*}

\begin{figure}
\begin{center}
\includegraphics[width=90mm]{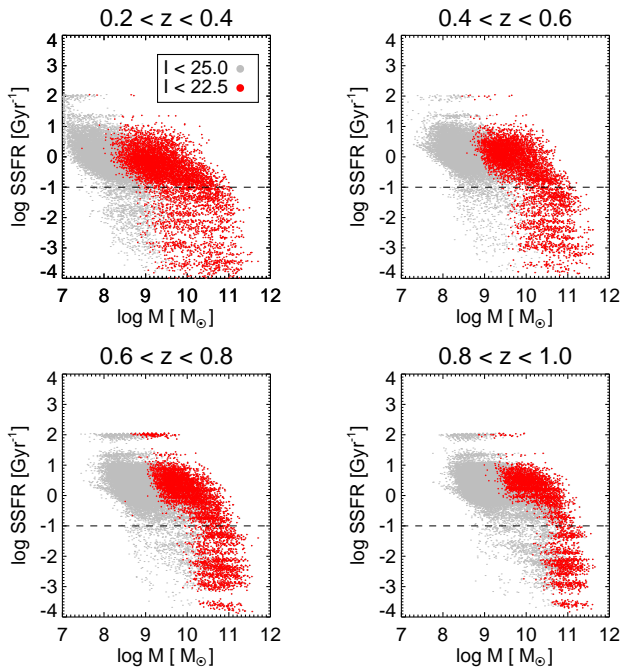} 
\caption{
SSFR vs. stellar mass for galaxies with $I_{\rm F814W} < 22.5$ (red dots) and 
for those with $I_{\rm F814W} < 25.0$ (grey dots) for each redshift bin.
}
\label{fig:msbias}
\end{center}
\end{figure}

We show the fraction of clumpy galaxies in star-forming galaxies with 
$SSFR > 0.1$ Gyr$^{-1}$ 
as a function of SFR in Figure \ref{fig:sfrac}. 
It is shown that the fraction of clumpy galaxies 
clearly increases with increasing SFR in all the redshift ranges. 
The fraction at a given SFR decreases with time over the wide range of SFR, 
although the strength of the evolution seems to depend on SFR. 
The fraction at $SFR \sim 10^{1.75} M_{\odot}$ yr$^{-1}$ decreases from 
$\sim 0.4$ at $0.8<z<1.0$ to $\sim0.05$ at $0.2<z<0.4$, while that at 
$SFR \sim 10^{0.75} M_{\odot}$ yr$^{-1}$ changes from $\sim 0.2$ 
to $\sim 0.07$ in the same redshift range. 
A slightly higher fraction around $SFR \sim 10^{1.75} M_{\odot}$ yr$^{-1}$ 
for galaxies at $0.8<z<1.0$ probably 
corresponds to that at $M_{\rm star} \sim 10^{10} M_{\odot}$ 
in Figure \ref{fig:mfrac} mentioned above.

Furthermore, we investigated the fraction of clumpy galaxies as a function of 
SSFR in each mass and redshift range. Figure \ref{fig:msfrac} shows 
the SSFR distribution and the fraction of clumpy galaxies
as a function of SSFR. In this figure, we used all sample galaxies, 
including galaxies with $SSFR < 0.1$ Gyr$^{-1}$. 
While the range of SSFRs of clumpy galaxies is similar to that 
of the other (non-clumpy) star-forming galaxies as shown in Figure \ref{fig:mstar}, 
the distribution of SSFR of clumpy galaxies tends to be skewed toward higher values.
In fact, the fraction of clumpy galaxies clearly 
increases with increasing SSFR in all the stellar mass 
and redshift ranges.  
Interestingly, we found that the fraction at a given SSFR 
is nearly independent of stellar mass in each redshift bin. 
In Figure \ref{fig:fssfr}, we show the fraction of clumpy galaxies as a function of 
SSFR for the different mass ranges in the same panel. 
The SSFR dependences of the fractions for the different stellar mass ranges are 
similar in each redshift bin, although the uncertainty of each data point is 
relatively large, especially at very high SSFR. 
In all the redshift ranges, 
we can see that the fraction increases with SSFR at $SSFR > 0.1$ Gyr$^{-1}$, 
while it is negligible at $SSFR < 0.1$ Gyr$^{-1}$. 
We can also see that the fraction of clumpy galaxies at a given SSFR decreases with time. 
For example, the fraction at $SSFR \sim 1$ Gyr$^{-1}$ changes from $\sim 0.3$ at 
$0.8<z<1.0$ to $\sim 0.07$ at $0.2<z<0.4$, while that at $SSFR \sim 10$ Gyr$^{-1}$ decreases from 
$\sim 0.4$ to $\sim 0.15$ in the same redshift range. 
We fit the data points at $SSFR > 0.1$ Gyr$^{-1}$ in each panel of Figure 
\ref{fig:fssfr} with a linear line of $f_{\rm clumpy} = a \times  
[ \log(SSFR) + 1.0 ]$, 
which satisfies $f_{\rm clumpy} = 0$ at $\log(SSFR)=-1$ 
(the dashed lines in the figure).  
The best-fit slope of the linear line on the $f_{\rm clumpy}$ -- 
$\log({\rm SSFR})$ plane 
clearly becomes steeper with redshift.  
Therefore, the fraction of clumpy galaxies 
at a given SSFR does not depend on stellar mass at each redshift,  
but evolves with time from $z\sim1$ to $z\sim0.2$.  

\begin{figure}
\begin{center}
\includegraphics[width=90mm]{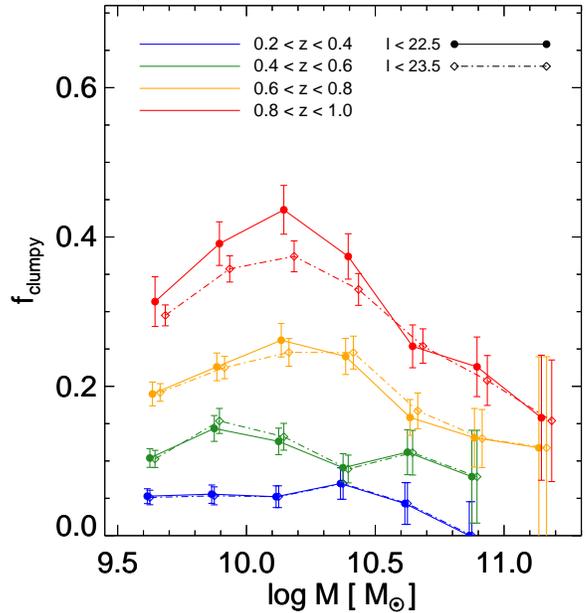} 
\caption{
The same as Figure \ref{fig:mfrac}, but the results for galaxies with 
$I_{\rm F814W} < 23.5$
are also shown (dashed-dotted lines)
as well as those for the basic sample with $I_{\rm F814W} < 22.5$ (solid lines). 
}
\label{fig:biasmf}
\end{center}
\end{figure}

\subsection{Possible Biases in Our Analysis}
\label{sec:bias}
In this section, we checked the effects of possible biases on our results in the 
previous subsections. First, we considered the effects of our magnitude limit of 
$I_{\rm F814W}<22.5$, which is set to ensure a secure selection of clumpy galaxies.
Since the observed $I_{\rm F814W}$ band samples the rest-frame $B$ band at 
$z\sim0.9$, the relatively bright $I_{\rm F814W}$-band magnitude limit 
can lead to a bias for galaxies with higher SSFRs at lower stellar mass.
Therefore we may preferentially miss low-mass galaxies with relatively low SSFRs 
at higher redshift.
In order to check this point in more detail, 
we show the SSFR-M$_{\rm star}$ diagram for galaxies 
with $I_{\rm F814W}<22.5$ and those with $I_{\rm F814W}<25$ in 
Figure \ref{fig:msbias}. It is shown that galaxies with relatively low SSFRs 
are missed by the $I_{\rm F814W}$-band magnitude limit at $M_{\rm star} \lesssim 
10^{10} M_{\odot}$ in the $0.8<z<1.0$ bin, while almost all galaxies 
with $SSFR > 0.1$ Gyr$^{-1}$ are picked up down to $M_{\rm star} \sim 10^{9.5} 
M_{\odot}$ at lower redshifts. 
In Figure \ref{fig:biasmf}, we show the fraction of clumpy galaxies in star-forming 
galaxies as in Figure \ref{fig:mfrac} but for galaxies with $I_{\rm F814W}<23.5$. 
It is shown that the relatively high fraction around $M_{\rm star} \sim 10^{10} M_{\odot}$ 
for galaxies at $0.8<z<1.0$ seen in Figure \ref{fig:mfrac} becomes lower, 
and the fraction is consistent with a constant value of $\sim 0.3$.
This is because we include low-mass galaxies with relatively low SSFRs at high redshift 
by using galaxies down to $I_{\rm F814W}=23.5$. Although we can mitigate the bias 
against low-mass galaxies with low SSFRs by including more faint galaxies into 
the sample, we note that the selection of clumpy galaxies becomes less secure 
for galaxies at $I_{\rm F814W} > 22.5$. On the other hand, the bias caused by the 
$I_{\rm F814W}$-band magnitude limit does not seem to affect the results in 
Figures \ref{fig:msfrac} and \ref{fig:fssfr}, because the fraction of clumpy galaxies 
{\it at a given SSFR} is not changed by this effect. 
In fact, we performed the same analysis by using the different $I_{\rm F814W}$-band 
magnitude limits and confirmed that the trends seen in Figures \ref{fig:msfrac} and 
\ref{fig:fssfr} do not depend on the magnitude limit. 

Second, we examined the effect of the morphological K-correction in our analysis.
Since we selected our clumpy galaxies in the $HST$/ACS $I_{\rm F814W}$-band images, 
the morphological selection was done at the rest-frame $R$ band at $z\sim0.3$, 
while galaxies at $z\sim0.9$ were classified at the rest-frame $B$ band.
If the clumps tend to be more conspicuous in shorter wavelengths (e.g., 
\citealp{elm09a}; \citealp{wuy12}; see also \citealp{guo12}), our ability to 
select clumpy galaxies could be weaker at lower redshifts. 
In order to check this effect, we performed the same selection of clumpy galaxies 
at $0.3<z<0.5$, using $HST$/ACS $V_{\rm F606W}$-band images, which correspond to 
the rest-frame $B$ band for these galaxies. The $V_{\rm F606W}$-band data 
were obtained in the CANDELS survey (\citealp{gro11}; \citealp{koe11}) 
and cover a field of $\sim 260$ arcmin$^2$ 
in the COSMOS field. Although the area is $\sim $3.6\% of the 1.64 deg$^2$ field of 
the $I_{\rm F814W}$-band data, we can roughly estimate the effect of the morphological
K-correction. For $\sim$ 100 star-forming galaxies with $SSFR > 0.1$ Gyr$^{-1}$ 
at $0.3<z<0.5$, 
we performed the same morphological analysis with the $V_{\rm F606W}$-band data 
and found that the fraction of clumpy galaxies increases from 
$\sim 0.1$ at the rest-frame $V$ band (the observed $I_{\rm F814}$ band) to 
$\sim$ 0.3--0.4 at the rest-frame $B$ band. \cite{wuy12} also pointed out 
that the morphological K-correction significantly affects the fraction of clumpy 
galaxies at $z\sim2$.  
Taking account of these results, we keep in mind the effect of 
the morphological K-correction when discussing the evolution of the fraction of 
clumpy galaxies in the following section.


\section{DISCUSSION}
\label{sec:dis}

In this study, we constructed a large sample of  
 clumpy galaxies at $0.2<z<1.0$ in the COSMOS field 
using the $HST$/ACS data and investigated the fraction of these galaxies and its 
evolution as a function of stellar mass, SFR, and SSFR.  
This is the first systematic search for clumpy galaxies at $z<1$.
Our main results are as follows. 
\begin{itemize}

\item The fraction of clumpy galaxies in star-forming galaxies decreases with time 
from $\sim 0.35$ at $0.8<z<1.0$ to $\sim 0.05$ at $0.2<z<0.4$ irrespective of stellar 
mass, although the fraction tends to be slightly lower at $M_{\rm star}>10^{10.5} 
M_{\odot}$ in each redshift bin.

\item The fraction of clumpy galaxies increases with increasing both SFR and SSFR in all the 
redshift ranges we investigated. In particular, the SSFR dependences of the fractions  
are similar among galaxies with different stellar masses. Moreover, the fraction at a given 
SSFR does not depend on stellar mass in each redshift bin. 

\item The fraction of clumpy galaxies at a given SSFR decreases with time at 
$SSFR > 0.1$ Gyr$^{-1}$. This can be explained by the effect of the 
morphological K-correction.
\end{itemize}

We discuss these results and their implications for both origins and evolution of clumpy 
galaxies in the following subsections.

\subsection{SSFR dependence of the fraction of clumpy galaxies} 
We found that the fraction of clumpy galaxies increases with increasing both SFR and SSFR. 
The similar fractions at a given SSFR among galaxies with different stellar masses 
may indicate that the SSFR is more important and fundamental physical parameter 
for the origin of the clumpy morphology. Among previous studies on clumpy galaxies, 
\cite{bou12} studied 14 clumpy galaxies and 13 smooth disk galaxies at $z\sim0.7$ 
selected by an eyeball classification, and found that the average and median 
SSFRs of clumpy galaxies are higher than those of smooth disk galaxies. 
\cite{sal12} also reported that clumpy galaxies at $0.5<z<1.3$ have 
systematically higher SSFRs than the other star-forming galaxies at the same 
redshifts. They selected clumpy galaxies with a quantitative clumpiness parameter, 
but their measurement of the clumpiness includes the surface brightness 
fluctuation on relatively small scales. The SSFR dependence of the fraction of 
clumpy galaxies seen in Figure \ref{fig:msfrac} is consistent with the results 
of these previous studies, although the selection methods for clumpy galaxies are 
different among the studies. 

Since the SSFR is  
 a current birth rate of stars relative to the integrated past star formation rate, 
it can be considered to represent the evolutionary stages of the stellar 
mass assembly by the star formation. In this view, 
the relatively high SSFRs of clumpy galaxies indicates that these galaxies may be  
systematically in younger stages in their star formation history.  
We can also consider the SSFR as a proxy for the gas mass to stellar mass 
ratio, M$_{\rm gas}$/M$_{\rm star}$, if we naively assume that the SFRs of galaxies  
roughly reflect their gas mass. 
Clumpy galaxies are expected to be (probably young) objects 
with relatively high gas mass fraction in this case. 
Recently, the gravitational instability and fragmentation in gas-rich disks 
is often proposed as a possible origin of the clumpy morphology of high-redshift galaxies, 
which show both coherent rotation and relatively large velocity dispersion in their gas 
(e.g., \citealp{imm04}; \citealp{bou07}; \citealp{dek09}; \citealp{bou10}; 
\citealp{gen12}). Gas-rich rotational disks are gravitationally 
unstable for the fragmentation and lead to the formation of large clumps. 
In this framework, the gas mass fraction is a key physical parameter. 
The stability for the gravitational fragmentation of the disks and the maximum 
unstable mass scale strongly depend on the gas mass fraction (e.g., \citealp{esc08}; 
\citealp{cac12}). If the SSFR is closely related to the gas mass fraction, 
the strong SSFR dependence of the fraction of clumpy galaxies in Figure \ref{fig:msfrac} 
can be explained by the relationship between the gravitational fragmentation and 
the gas mass fraction of the rotational disks.

\cite{esc11} also discussed that a large maximum unstable mass of gas-rich disks 
corresponds to a large velocity dispersion of turbulent motions of gas 
in the self-regulated quasi-stationary state with the Toomre parameter $Q\sim1$. 
They also claimed
 that the large velocity dispersion can cause an enhancement of star formation 
activity. In fact, the correlation between the mass of the most massive clump of 
galaxies and their surface SFR density has been observed at $z\gtrsim 1$ 
(\citealp{liv12}; \citealp{swi12}). This scenario may explain the relatively 
higher SSFRs of clumpy galaxies. 
 
Another possible origin of the clumpy morphology is the galaxy merger 
(e.g., \citealp{som01}; \citealp{lot04}). 
For example, \cite{dim08} performed extensive numerical simulations of major mergers 
and found that gas-rich major mergers can cause the greater disk fragmentation 
than the cases of isolated gas-rich galaxies. 
Morphological 
studies of high-redshift galaxies and comparisons of these objects 
with merger galaxies in 
the nearby universe suggested that some fraction of clumpy and irregular galaxies 
at $z\gtrsim 1$ are ongoing mergers (e.g., \citealp{lot08}; \citealp{pet09}; 
\citealp{ove10}).
\cite{man13} suggested that a non-negligible fraction of large clumps in 
high-redshift clumpy galaxies come from minor mergers, based on a numerical 
cosmological simulation of disk galaxies.
\cite{pue10} reported that a significant fraction of intermediate-mass 
clumpy galaxies at $z\sim0.6$ have complex kinematics, which is compatible with 
major mergers.
In this scenario, the SSFR dependence of the fraction of clumpy galaxies can be 
understood by the enhancement of star formation caused by the galaxy mergers.
The galaxy interaction/merger does not only cause disturbed and clumpy morphologies, 
but also triggers intense star formation. Thus the SSFRs of clumpy galaxies tend to be 
enhanced from the main sequence of star-forming galaxies. 
However, it is unclear whether 
such starbursts by mergers are consistent with the relatively tight SFR-M$_{\rm star}$ 
relation or not (e.g., \citealp{noe07}; \citealp{elb07}; \citealp{ren09}; 
\citealp{wuy11}; \citealp{rod11}).

\begin{figure}
\begin{center}
\includegraphics[width=90mm]{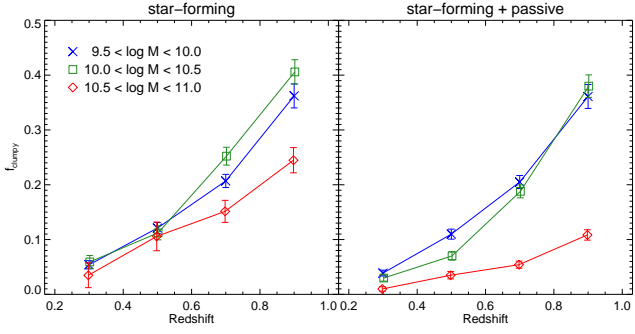} 
\caption{{\bf left:} 
Evolution of the fraction of clumpy galaxies in star-forming galaxies 
($SSFR > 0.1$ Gyr$^{-1}$) with 
different stellar masses. {\bf right:} The same as the left panel but for 
all galaxies with $I_{\rm F814W} < 22.5$ including passively evolving galaxies.
}
\label{fig:down}
\end{center}
\end{figure}

\subsection{Evolution of the fraction of clumpy galaxies}
We found that the overall fraction of clumpy galaxies in star-forming galaxies 
with $M_{\rm star} > 10^{9.5} M_{\odot}$ 
decreases from $\sim 0.35$ at $z\sim0.9$ to $\sim 0.05$ at $z\sim0.3$. 
While many actively star-forming galaxies with clumpy morphologies have been observed
at $z>1$ (e.g., \citealp{elm07}; 
\citealp{gen08}; \citealp{for11}), most relatively bright galaxies belong to the Hubble 
sequence and clumpy galaxies are very rare at $z\sim0$ (e.g., \citealp{ove10}). 
Our result naturally connects between these previous studies in the early universe and 
those in the nearby universe, although the morphological K-correction may  
affect the result (Section \ref{sec:bias}). 
\cite{wuy12} reported that the fraction of clumpy galaxies in star-forming galaxies at 
$z\sim1$ is 27 \% when the selection for clumps was performed at the rest-frame $V$ band. 
Our result at $0.8<z<1.0$ ($\sim 30$\%), 
which was obtained at the rest-frame $B$ band (see Section \ref{sec:bias}), 
is consistent with their result, although the selection criteria for clumpy galaxies 
are different. 

Several authors pointed out that the clumpy morphology persists to lower redshifts 
in lower-mass galaxies, so called ``down-sizing effect'' in the clumpy morphology 
(e.g., \citealp{elm09b}; \citealp{elm11}). There are many bright/massive clumpy 
galaxies at $z\gtrsim1$, while such clumpy morphology can be seen only in 
low-mass systems such as dwarf irregular galaxies in the present universe. 
In Figure \ref{fig:down}, we show the evolution of the fraction of clumpy galaxies 
with different stellar masses in order to investigate the down-sizing effect.
The evolution is similar among the different 
mass samples, while the fraction of clumpy galaxies is slightly lower 
in massive galaxies with $M_{\rm star} > 10^{10.5} M_{\odot}$. 
The differences of the fraction among galaxies with different masses become 
larger when we include passive galaxies with $SSFR < 0.1$ Gyr$^{-1}$ 
into the samples 
(the right panel of Figure \ref{fig:down}). 
If we extrapolate the trend in Figure \ref{fig:down} to the present,  
the fraction of clumpy galaxies in more massive galaxies is expected to become 
negligible earlier. Therefore, our result is not inconsistent with 
the down-sizing picture, but the mass dependence of the evolution is weak 
in the stellar mass range we investigated.
The lower fraction of clumpy galaxies in massive star-forming galaxies
with $M_{\rm star} > 10^{10.5}M_{\odot}$ at each redshift 
may be explained by lower SSFR of these massive galaxies (Figure \ref{fig:msbias}). 

\begin{figure}
\begin{center}
\includegraphics[width=90mm]{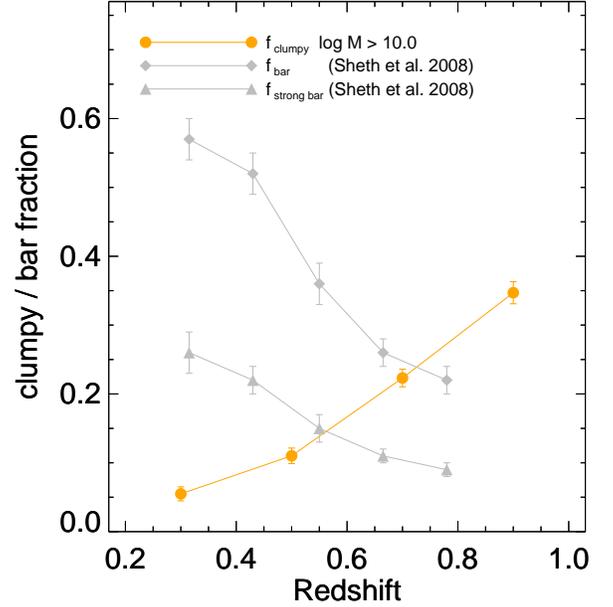} 
\caption{
Fraction of clumpy galaxies in star-forming galaxies
 with $M_{\rm star} > 10^{10} M_{\odot}$ 
and the fractions of barred spiral galaxies from 
\cite{she08} as a function of redshift.
Diamonds show the fraction of all barred spiral galaxies, while 
triangles represent that of strong barred galaxies. 
These fractions of barred galaxies 
are measured in face-on spiral galaxies with inclination angles of 
 $i > 65^\circ$ (see \citealp{she08} for details).  
}
\label{fig:bar}
\end{center}
\end{figure}

Another our result of the SSFR dependence of the fraction of clumpy galaxies 
seen in all the mass and redshift ranges indicates that the evolution of the SSFRs 
of galaxies leads to the evolution of the fraction of clumpy galaxies.
In fact, the median SSFR of star-forming galaxies decreases by $\sim 1$ dex 
from $z\sim 0.9$ to $z\sim 0.3$ (Figure \ref{fig:msfrac}). 
For example, if we assume the relation between the fraction of clumpy galaxies 
and SSFR at $0.8<z<1.0$ shown in Figure \ref{fig:fssfr}, the decrease of the SSFR 
from $SSFR \sim 10^{0.25}$ Gyr$^{-1}$ (median value at $z\sim0.9$) to 
$SSFR \sim 10^{-0.75}$ Gyr$^{-1}$ (that at $z\sim0.3$) corresponds to the 
evolution of the fraction from $\sim 0.35$ to $\sim 0.05$. 
Thus the evolution of the fraction of clumpy galaxies in star-forming galaxies 
at $0.2<z<1.0$ appear to be explained by the evolution of the SSFR. 
On the other hand, from such correlation between the fraction of clumpy galaxies and 
SSFR, the fraction of clumpy galaxies is expected to be higher at 
higher redshifts, because galaxies tend to have higher SSFRs than those at $z\lesssim1$.
\cite{wuy12} found that the fraction of clumpy galaxies in star-forming galaxies 
with $M_{\rm star} > 10^{10} M_{\odot}$ at $1.5<z<2.5$ is 42\% when the morphological 
selection was done at the rest-frame $V$ band. \cite{tad14} also reported that 
42\% of H$\alpha$ emitters at $z\sim 2.2$ and 2.5 have clumpy morphology, 
although their clump selection was performed with both rest-frame UV and optical-bands 
images. The average SSFRs of star-forming galaxies at $z\sim2$ in both studies 
are $\sim 10^{0.5}$ Gyr$^{-1}$, and therefore the fractions of clumpy galaxies 
of $\sim 40$\% in these studies seem to be consistent with the relation between 
the fraction of clumpy galaxies and SSFR at $0.8<z<1.0$ shown in Figure \ref{fig:fssfr}.

In the gravitational fragmentation model for the formation of giant clumps in disk 
galaxies, the rapid and smooth streams of gas along filaments effectively 
penetrate halos of galaxies at high redshift (e.g., \citealp{ker05}; \citealp{dek09b}). 
This ``cold accretion'' keeps active star formation and a high gas mass fraction of 
these galaxies, which leads to the formation of the clumpy morphology.  
In fact, such high gas mass fraction of star-forming galaxies at $z\sim2$ have been 
observed (e.g., \citealp{dad10}; \citealp{tac13}), and the observed high turbulent 
velocity of gas in high-redshift clumpy galaxies also supports this scenario 
(e.g., \citealp{cre09}; \citealp{for09}). Using an analytic model, 
\cite{cac12} predicted that such disks tend to stabilize at $z \lesssim 1$
 mainly due to the decrease of the gas mass fraction. They suggested that the decrease is 
attributed to   
the gradual decline of the cosmological accretion rate into halos of galaxies with time 
(e.g., \citealp{gen08}),  the gas consumption by the star formation,  the inflows of 
clumps into the center of galaxies by the gravitational torque 
(e.g, \citealp{dek09}), and the gas outflows by the supernova feedback. 
If the SSFRs of galaxies are closely related to the gas mass fraction as 
discussed above, the evolution of the fraction of clumpy galaxies at $0.2<z<1.0$ 
 could be explained by this scenario. The decrease of the gas mass fraction with time 
at $z\lesssim 1$ 
causes the stabilization of galactic disks, while it also leads to  
the decrease of the SSFRs of these galaxies. 
Interestingly, several studies reported that the fraction of barred spiral galaxies 
increases with time in the same redshift range (e.g., \citealp{abr99}; 
 \citealp{she08}; \citealp{mel14}). 
The bar instability is considered to occur in ``mature''systems where 
stellar disk is dynamically cold and rotationally supported, and the surface stellar 
density is sufficiently high. Clumpy galaxies may evolve to these barred galaxies 
when the gas fraction becomes lower and the stellar disk is stabilized (e.g., 
\citealp{imm04}; \citealp{she12}; \citealp{kra12}). 
In Figure \ref{fig:bar}, we compare the fraction of clumpy galaxies with 
the fraction of barred spiral galaxies in the same COSMOS field from \cite{she08}. 
Note that since the bar fraction in \cite{she08} is the fraction of barred galaxies 
in face-on spiral galaxies with $i > 65^\circ$ excluding irregular galaxies, 
it is difficult to directly compare the absolute values of the both fractions.
Nevertheless, we can see the fraction of barred spiral galaxies increases with 
time, as the fraction of clumpy galaxies decreases in the figure.
The transition from clumpy galaxies to barred spiral galaxies may gradually 
occur from $z\sim1$ to $z\sim0$.


\begin{figure}
\begin{center}
\includegraphics[width=90mm]{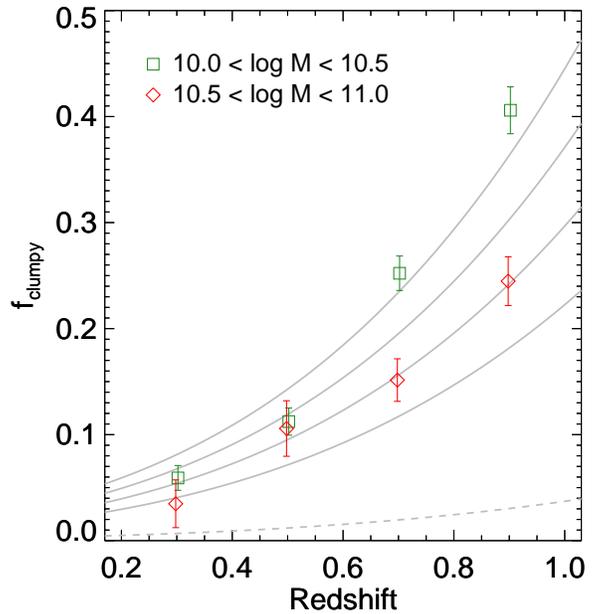} 
\caption{
Comparison between the observed fraction of clumpy galaxies 
 and that expected 
from wet major merger rate by \cite{lop13}. The solid lines show 
the major merger rate multiplied by a time scale of 3, 4, 5, and 6 Gyr, while 
the dashed line represents that multiplied by a time scale of 0.5 Gyr. 
}
\label{fig:merger}
\end{center}
\end{figure}

On the other hand, in the major merger scenario for the origin of 
the clumpy morphology, the evolution of the major merger 
rate may explain the evolution of the fraction of clumpy galaxies at $0.2<z<1.0$. 
Many observational studies have found that the major merger rate decreases with 
time at $z\lesssim 1$ (e.g., \citealp{lef00}; \citealp{con03}; \citealp{lin08}; 
\citealp{der09}; \citealp{xu12}; see also \citealp{lot11}).
In particular, several studies reported that 
the gas-rich wet major merger, which is considered to be important 
for making the clumpy morphology \citep{dim08}, decreases with time in the 
redshift range (\citealp{cho11}; \citealp{pue12}; \citealp{lop13}).
Figure \ref{fig:merger} compares the observed fraction of clumpy galaxies at 
$0.2<z<1.0$ with that expected from the wet major merger rate 
of galaxies with $M_{\rm star} \sim 10^{10}-10^{10.5} M_{\odot}$ by \cite{lop13}.
We multiplied the major merger rate from \cite{lop13} by an arbitrary time scale  
when the merged galaxies are seen as clumpy galaxies to estimate the expected 
fraction of clumpy galaxies. The evolution of clumpy galaxies can be roughly 
explained by the evolution of the major merger rate with 
the time scales of $\sim$ 3--6 Gyr. However, these time scales seem to be 
too long for the merger time scale during which the morphology is disturbed and 
clumpy (e.g., \citealp{dim08}; \citealp{lot10}). 
If we assume a typical merger time scale of 
$\sim 0.5$ Gyr, the expected fraction of clumpy galaxies become much lower than 
the observed fraction at $z\gtrsim 0.5$ (the dashed line in Figure \ref{fig:merger}).
It seems to be difficult to explain the fraction of clumpy galaxies at $0.2<z<1.0$  
only by the wet major merger.

Finally, we note that the fraction of clumpy galaxies at a given SSFR decreases with time
from $z\sim 0.9$ to $z\sim0.3$. This can be due to the morphological K-correction  
because we selected clumpy galaxies at the observed $I_{\rm F814W}$ band as 
discussed in Section \ref{sec:bias}. If this is the case, the intrinsic fraction of 
clumpy galaxies at a given SSFR could not depend on redshift. There may be 
the universal relation between the fraction of clumpy galaxies and SSFR.

\vspace{1pc}

We would like to thank the anonymous referee for valuable comments and suggestions. 
We also thank Tsutomu T. Takeuchi at Nagoya University for his generous
support to K. L. M. and for useful discussion.
The HST COSMOS Treasury program was supported through NASA grant
HST-GO-09822. We greatly acknowledge the contributions of the entire
COSMOS collaboration consisting of more than 70 scientists. 
This work was financially supported in part by the Japan Society for
the Promotion of Science (Nos. 17253001, 19340046, 23244031, and 23740152).


{}




\clearpage

\clearpage

\clearpage

\clearpage


\begin{thebibliography}{}

\bibitem[Abraham et al.(1999)]{abr99} Abraham, R.~G., 
Merrifield, M.~R., Ellis, R.~S., Tanvir, N.~R., 
\& Brinchmann, J.\ 1999, \mnras, 308, 569 

\bibitem[Bertin 
\& Arnouts(1996)]{ber96} Bertin, E., \& Arnouts, S.\ 1996, \aaps, 117, 393 

\bibitem[Bournaud et al.(2007)]{bou07} Bournaud, F., 
Elmegreen, B.~G., \& Elmegreen, D.~M.\ 2007, \apj, 670, 237 

\bibitem[Bournaud et al.(2010)]{bou10} Bournaud, F., 
Elmegreen, B.~G., Teyssier, R., Block, D.~L., 
\& Puerari, I.\ 2010, \mnras, 409, 1088 


\bibitem[Bournaud et al.(2012)]{bou12} Bournaud, F., Juneau, 
S., Le Floc'h, E., et al.\ 2012, \apj, 757, 81
\bibitem[Bruzual 
\& Charlot(2003)]{bru03} Bruzual, G., \& Charlot, S.\ 2003, \mnras, 344, 1000 

\bibitem[Cacciato et al.(2012)]{cac12} Cacciato, M., Dekel, 
A., \& Genel, S.\ 2012, \mnras, 421, 818 



\bibitem[Calzetti et al.(2000)]{cal00} Calzetti, D., Armus, 
L., Bohlin, R.~C., et al.\ 2000, \apj, 533, 682

\bibitem[Cameron et al.(2011)]{cam11} Cameron, E., Carollo, 
C.~M., Oesch, P.~A., et al.\ 2011, \apj, 743, 146 

\bibitem[Chabrier(2003)]{cha03} Chabrier, G.\ 2003, \pasp, 
115, 763

\bibitem[Chou et al.(2011)]{cho11} Chou, R.~C.~Y., Bridge, 
C.~R., \& Abraham, R.~G.\ 2011, \aj, 141, 87


\bibitem[Conselice et al.(2003)]{con03} Conselice, C.~J., 
Bershady, M.~A., Dickinson, M., \& Papovich, C.\ 2003, \aj, 126, 1183

\bibitem[Cowie et al.(1995)]{cow95} Cowie, L.~L., Hu, E.~M., 
\& Songaila, A.\ 1995, \aj, 110, 1576

\bibitem[Cresci et al.(2009)]{cre09} Cresci, G., Hicks, 
E.~K.~S., Genzel, R., et al.\ 2009, \apj, 697, 115 

\bibitem[Daddi et al.(2010)]{dad10} Daddi, E., Bournaud, F., 
Walter, F., et al.\ 2010, \apj, 713, 686

\bibitem[Dekel et al.(2009a)]{dek09} Dekel, A., Sari, R., 
\& Ceverino, D.\ 2009a, \apj, 703, 785

\bibitem[Dekel et al.(2009b)]{dek09b} Dekel, A., Birnboim, Y., 
Engel, G., et al.\ 2009b, \nat, 457, 451

\bibitem[de Ravel et 
al.(2009)]{der09} de Ravel, L., Le F{\`e}vre, O., Tresse, L., et al.\ 2009, \aap, 498, 379 

\bibitem[Di Matteo et 
al.(2008)]{dim08} Di Matteo, P., Bournaud, F., Martig, M., et al.\ 2008, \aap, 492, 31

\bibitem[Elbaz et al.(2007)]{elb07} Elbaz, D., Daddi, E., Le Borgne, D., et al.\ 2007, \aap, 468, 33 

\bibitem[Elmegreen(2011)]{elm11} Elmegreen, B.~G.\ 2011, EAS 
Publications Series, 51, 59

\bibitem[Elmegreen et al.(2009a)]{elm09a} Elmegreen, B.~G., 
Elmegreen, D.~M., Fernandez, M.~X., \& Lemonias, J.~J.\ 2009a, \apj, 692, 12

\bibitem[Elmegreen et al.(2009b)]{elm09b} Elmegreen, D.~M., 
Elmegreen, B.~G., Marcus, M.~T., et al.\ 2009b, \apj, 701, 306 


\bibitem[Elmegreen et al.(2007)]{elm07} Elmegreen, D.~M., 
Elmegreen, B.~G., Ferguson, T., \& Mullan, B.\ 2007, \apj, 663, 734 

\bibitem[Elvis et al.(2009)]{elv09} Elvis, M., Civano, F., 
Vignali, C., et al.\ 2009, \apjs, 184, 158 

\bibitem[Escala 
\& Larson(2008)]{esc08} Escala, A., \& Larson, R.~B.\ 2008, \apjl, 685, L31

\bibitem[Escala(2011)]{esc11} Escala, A.\ 2011, \apj, 735, 56 

\bibitem[F{\"o}rster Schreiber et al.(2006)]{for06} 
F{\"o}rster Schreiber, N.~M., Genzel, R., Lehnert, M.~D., et al.\ 2006, 
\apj, 645, 1062

\bibitem[F{\"o}rster Schreiber et al.(2009)]{for09} 
F{\"o}rster Schreiber, N.~M., Genzel, R., Bouch{\'e}, N., et al.\ 2009, 
\apj, 706, 1364 

\bibitem[F{\"o}rster Schreiber et al.(2011)]{for11} 
F{\"o}rster Schreiber, N.~M., Shapley, A.~E., Genzel, R., et al.\ 2011, 
\apj, 739, 45


\bibitem[Genzel et al.(2008)]{gen08} Genzel, R., Burkert, A., 
Bouch{\'e}, N., et al.\ 2008, \apj, 687, 59 


\bibitem[Genzel et al.(2012)]{gen12} Genzel, R., Tacconi, 
L.~J., Combes, F., et al.\ 2012, \apj, 746, 69

\bibitem[Grogin et al.(2011)]{gro11} Grogin, N.~A., Kocevski, 
D.~D., Faber, S.~M., et al.\ 2011, \apjs, 197, 35

\bibitem[Guo et al.(2012)]{guo12} Guo, Y., Giavalisco, M., 
Ferguson, H.~C., Cassata, P., \& Koekemoer, A.~M.\ 2012, \apj, 757, 120 

\bibitem[Hasinger et al.(2007)]{has07} Hasinger, G., 
Cappelluti, N., Brunner, H., et al.\ 2007, \apjs, 172, 29

\bibitem[Hubble(1936)]{hub36} Hubble, E.~P.\ 1936, Realm of 
the Nebulae, by E.P.~Hubble.~ New Haven: Yale University Press, 1936.~ ISBN 
9780300025002, 

\bibitem[Ilbert et 
al.(2013)]{ilb13} Ilbert, O., McCracken, H.~J., Le F{\`e}vre, O., et al.\ 2013, \aap, 556, A55 q


\bibitem[Ilbert et al.(2010)]{ilb10} Ilbert, O., Salvato, M., 
Le Floc'h, E., et al.\ 2010, \apj, 709, 644 


\bibitem[Ilbert et al.(2009)]{ilb09} Ilbert, O., Capak, P., 
Salvato, M., et al.\ 2009, \apj, 690, 1236 

\bibitem[Immeli et al.(2004)]{imm04} Immeli, A., Samland, M., 
Westera, P., \& Gerhard, O.\ 2004, \apj, 611, 20 

\bibitem[Kajisawa 
\& Yamada(2001)]{kaj01} Kajisawa, M., \& Yamada, T.\ 2001, \pasj, 53, 833 

\bibitem[Kajisawa et al.(2010)]{kaj10} Kajisawa, M., 
Ichikawa, T., Yamada, T., et al.\ 2010, \apj, 723, 129 

\bibitem[Kere{\v s} et al.(2005)]{ker05} Kere{\v s}, D., 
Katz, N., Weinberg, D.~H., \& Dav{\'e}, R.\ 2005, \mnras, 363, 2

\bibitem[Koekemoer et al.(2007)]{koe07} Koekemoer, A.~M., 
Aussel, H., Calzetti, D., et al.\ 2007, \apjs, 172, 196 

\bibitem[Koekemoer et al.(2011)]{koe11} Koekemoer, A.~M., 
Faber, S.~M., Ferguson, H.~C., et al.\ 2011, \apjs, 197, 36

\bibitem[Kraljic et al.(2012)]{kra12} Kraljic, K., Bournaud, 
F., \& Martig, M.\ 2012, \apj, 757, 60


\bibitem[Le F{\`e}vre et al.(2000)]{lef00} Le F{\`e}vre, O., 
Abraham, R., Lilly, S.~J., et al.\ 2000, \mnras, 311, 565 

\bibitem[Lin et al.(2008)]{lin08} Lin, L., Patton, D.~R., 
Koo, D.~C., et al.\ 2008, \apj, 681, 232

\bibitem[Livermore et al.(2012)]{liv12} Livermore, R.~C., 
Jones, T., Richard, J., et al.\ 2012, \mnras, 427, 688 

\bibitem[L{\'o}pez-Sanjuan et 
al.(2013)]{lop13} L{\'o}pez-Sanjuan, C., Le F{\`e}vre, O., Tasca, L.~A.~M., et al.\ 2013, \aap, 553, A78

\bibitem[Lotz et al.(2004)]{lot04} Lotz, J.~M., Primack, J., 
\& Madau, P.\ 2004, \aj, 128, 163 

\bibitem[Lotz et al.(2008)]{lot08} Lotz, J.~M., Davis, M., 
Faber, S.~M., et al.\ 2008, \apj, 672, 177 

\bibitem[Lotz et al.(2010)]{lot10} Lotz, J.~M., Jonsson, P., 
Cox, T.~J., \& Primack, J.~R.\ 2010, \mnras, 404, 590 

\bibitem[Lotz et al.(2011)]{lot11} Lotz, J.~M., Jonsson, P., 
Cox, T.~J., et al.\ 2011, \apj, 742, 103

\bibitem[Mandelker et al.(2013)]{man13} Mandelker, N., Dekel, 
A., Ceverino, D., et al.\ 2013, arXiv:1311.0013

\bibitem[Melvin et al.(2014)]{mel14} Melvin, T., Masters, K., 
Lintott, C., et al.\ 2014, \mnras, 438, 2882

\bibitem[Noeske et al.(2007)]{noe07} Noeske, K.~G., Weiner, 
B.~J., Faber, S.~M., et al.\ 2007, \apjl, 660, L43 

\bibitem[Noguchi(1998)]{nog98} Noguchi, M.\ 1998, \nat, 392, 
253 


\bibitem[Overzier et al.(2009)]{ove09} Overzier, R.~A., 
Heckman, T.~M., Tremonti, C., et al.\ 2009, \apj, 706, 203 


\bibitem[Overzier et al.(2010)]{ove10} Overzier, R.~A., 
Heckman, T.~M., Schiminovich, D., et al.\ 2010, \apj, 710, 979 


\bibitem[Petty et al.(2009)]{pet09} Petty, S.~M., de Mello, 
D.~F., Gallagher, J.~S., III, et al.\ 2009, \aj, 138, 362

\bibitem[Puech(2010)]{pue10} Puech, M.\ 2010, \mnras, 406, 
535 

\bibitem[Puech et al.(2012)]{pue12} Puech, M., Hammer, F., 
Hopkins, P.~F., et al.\ 2012, \apj, 753, 128

\bibitem[Renzini(2009)]{ren09} Renzini, A.\ 2009, \mnras, 
398, L58 

\bibitem[Rodighiero et al.(2011)]{rod11} Rodighiero, G., 
Daddi, E., Baronchelli, I., et al.\ 2011, \apjl, 739, L40 

\bibitem[Salmi et al.(2012)]{sal12} Salmi, F., Daddi, E., 
Elbaz, D., et al.\ 2012, \apjl, 754, L14 


\bibitem[Santini et 
al.(2009)]{san09} Santini, P., Fontana, A., Grazian, A., et al.\ 2009, \aap, 504, 751 


\bibitem[Scoville et al.(2007)]{sco07} Scoville, N., Aussel, 
H., Brusa, M., et al.\ 2007, \apjs, 172, 1 

\bibitem[Sheth et al.(2008)]{she08} Sheth, K., Elmegreen, 
D.~M., Elmegreen, B.~G., et al.\ 2008, \apj, 675, 1141

\bibitem[Sheth et al.(2012)]{she12} Sheth, K., Melbourne, J., 
Elmegreen, D.~M., et al.\ 2012, \apj, 758, 136 

\bibitem[Somerville et al.(2001)]{som01} Somerville, R.~S., 
Primack, J.~R., \& Faber, S.~M.\ 2001, \mnras, 320, 504

\bibitem[Steidel et al.(1996)]{ste96} Steidel, C.~C., 
Giavalisco, M., Dickinson, M., \& Adelberger, K.~L.\ 1996, \aj, 112, 352 

\bibitem[Swinbank et al.(2012)]{swi12} Swinbank, A.~M., 
Smail, I., Sobral, D., et al.\ 2012, \apj, 760, 130

\bibitem[Tacconi et al.(2010)]{tac10} Tacconi, L.~J., Genzel, 
R., Neri, R., et al.\ 2010, \nat, 463, 781

\bibitem[Tacconi et al.(2013)]{tac13} Tacconi, L.~J., Neri, 
R., Genzel, R., et al.\ 2013, \apj, 768, 74

\bibitem[Tadaki et al.(2014)]{tad14} Tadaki, K.-i., Kodama, 
T., Tanaka, I., et al.\ 2014, \apj, 780, 77

\bibitem[Wright et al.(2007)]{wri07} Wright, S.~A., Larkin, 
J.~E., Barczys, M., et al.\ 2007, \apj, 658, 78

\bibitem[Wuyts et al.(2011)]{wuy11} Wuyts, S., F{\"o}rster 
Schreiber, N.~M., Lutz, D., et al.\ 2011, \apj, 738, 106 

\bibitem[Wuyts et al.(2012)]{wuy12} Wuyts, S., F{\"o}rster 
Schreiber, N.~M., Genzel, R., et al.\ 2012, \apj, 753, 114 

\bibitem[Xu et al.(2012)]{xu12} Xu, C.~K., Shupe, D.~L., 
B{\'e}thermin, M., et al.\ 2012, \apj, 760, 72


\end{thebibliography}
\end{document}